\documentclass[english]{article}%
\usepackage{amssymb}
\usepackage{fontenc}
\usepackage[latin1]{inputenc}
\usepackage{amsmath}
\usepackage{fontenc}
\usepackage{babel}
\usepackage{amsfonts}
\usepackage{graphicx}%
\providecommand{\U}[1]{\protect\rule{.1in}{.1in}}
\begin{document}

\title{\textbf{Creation of neutral fermions with anomalous magnetic moments from a
vacuum by inhomogeneous magnetic fields}}
\author{S.P. Gavrilov\thanks{Department of General and Experimental Physics, Herzen
State Pedagogical University of Russia, Moyka embankment 48, 191186 St.
Petersburg, Russia; e-mail: gavrilovsergeyp@yahoo.com} and D.M. Gitman
\thanks{Institute of Physics, University of São Paulo, CP 66318, CEP 05315-970
São Paulo, SP, Brazil; e-mail: gitman@if.usp.br} }
\maketitle

\begin{abstract}
A consistent nonperturbative approach (based on QFT) to neutral fermion
creation (due to their magnetic moments) in strong inhomogeneous magnetic
fields is considered. It is demonstrated that quantization in terms of neutral
particles and antiparticles is possible in terms of the states with
well-defined spin polarization. Such states are localizable and can form wave
packets in a given asymptotic region. In this case, the problem can be
technically reduced to the problem of charged-particle creation by an electric
step. In particular, the relation to the Schwinger method of an effective
action is established. As an example, we calculate neutral fermion creation
from the vacuum by a linearly growing magnetic field. We show that the total
number and the vacuum-to-vacuum transition probability of created pairs depend
only on the gradient of the magnetic field, but not on its strength, and this
fact does not depend on the spacetime dimension.\emph{\ }We show\emph{\ }that
the created flux aimed in one of the directions is formed from fluxes of
particles and antiparticles of equal intensity and with the same magnetic
moments parallel to the external field. In such a flux, particle and
antiparticle velocities that are perpendicular to the plane of the magnetic
moment and flux direction are essentially depressed. The creation of neutral
fermions with anomalous magnetic moments leads to a smoothing of the initial
magnetic field, which in turn prevents appearance of superstrong constant
magnetic fields. Our estimations show that the vacuum instability with respect
to the creation of neutrinos and even neutrons in strong magnetic fields of
the magnetars and fields generated during a supernova explosion has to be
taken into account in the astrophysics. In particular, it may be of
significance for dark matter studies.

PACS numbers:12.20.Ds, 13.40.Em, 95.30.Cq,95.35.+d

Keywords: nonperturbative vacuum pair production, anomalous magnetic moment,
strong inhomogeneous magnetic field, dark matter

\end{abstract}

\section{Introduction\label{S1}}

Usually, particle creation from the vacuum by strong electromagnetic fields is
associated with the creation of charged particles by strong electric-like
fields. Acting on virtual charged particles, an electric-like field can
produce a work and materialize them on the mass shell as real particles.
Nevertheless, if a neutral particle has an anomalous magnetic moment, an
inhomogeneous magnetic field acting on such a particle, can also change its
kinetic energy (produce a work). This mechanism can provide neutral-particle
creation from the vacuum by strong inhomogeneous magnetic fields. In this
respect, one can speak about two candidates among the known elementary
particles: neutrons and neutrinos. It is known that the neutron has negative
magnetic moment given by $\mu_{n}=-1.9130427(5)\mu_{N}$, where $\mu_{N}$ is
the nuclear magneton, $\mu_{N}=e/2m_{N}$. It is also possible that neutrinos
have magnetic moments (in general, effective magnetic moments which take into
account neutrino mixing and the oscillations) acquired through quantum loop
effects; for a review see Refs. \cite{rev-nuMM2010,Dvo2010,rev-nu-em2012}.The
recent experimental constraints on the neutrino magnetic moments are in the
range $\sim10^{-11}\mu_{B}$ ($\mu_{\nu}<2.9\times10^{-11}\mu_{B}$ for electron
neutrino) \cite{exp-nuMM2010}, where $\mu_{B}=e/2m_{e}$ is the Bohr magneton.
Astrophysical constraints on the magnetic moment of the Dirac neutrino can be
even stronger, $\mu_{\nu}<1.1\times10^{-12}\mu_{B}$ \cite{astrMM2009}. Note
that in order to satisfy $m_{\nu}\lesssim1\;\mathrm{eV}$, the theory argues
that a more natural scale for the Dirac neutrino would be $\mu_{\nu}$
$\lesssim10^{-14}\mu_{B}$ \cite{Bell05}.

The discovery of neutrino masses suggests the likely existence of the light
sterile neutrinos that appear in the low-energy effective theory in most
extensions of the standard model, and in principle can have any mass, in
particular, in the $1\;\mathrm{eV}$ mass range. The sterile neutrinos with
masses of several keV can account for cosmological dark matter, e.g., see
Refs. \cite{Kuz09,LSnu12} for a recent review, and references therein. It is
possible that due to some new physics that the neutrino magnetic moment is
big. Various observational constraints on the magnetic moment $\mu$ of a dark
matter particle for masses $M$ in the range $1\;\mathrm{keV}$ to
$100\;\mathrm{MeV}$ have been considered in Refs. \cite{Sig-etal04,Gard09}.
The strongest limits on $\mu$ emerge at the lightest mass scales. For example,
if $M=m_{e}/10$ then $\left\vert \mu\right\vert <3.4\times10^{-5}\mu_{B}$ due
to precision electroweak measurements. It is noted \cite{Gard09} that a
variety of astrophysical constraints can be significantly weakened by the
candidate particle's mass and the above-mentioned constraints can be weakened
by other means as well.

The effect under discussion can be observed\emph{\ }in inhomogeneous magnetic
fields that have to be very strong in a certain domain\emph{.} Such fields can
exist in nature. It has been suggested that magnetic fields of order
$10^{15}-10^{16}\;\mathrm{G}$ or stronger, up to $10^{18}\;\mathrm{G}$, can
probably be generated during a supernova explosion or in the vicinity of the
special group of neutron stars know as magnetars, see; for example, Ref.
\cite{nstarB2003}. For magnetar cores made of quark matter the interior field
can be estimated to reach values $B\sim10^{20}\;\mathrm{G}$
\cite{quark-core11}. The possibility to create a strong quasiuniform magnetic
field with the strength of the hadronic scale $B\sim10^{19}\;\mathrm{G}$---or
even higher in heavy-ion collisions at RHIC and LHC, when the matter in the
central region is presumably in the quark-gluon plasma phase---was recently
shown \cite{QGPmagn09}. Superconducting cosmic string---if they exist---could
generate fields more then $10^{30}\;\mathrm{G}$ in their vicinities
\cite{cosmstringfield}.

Recently, the Schwinger effective action approach \cite{S51} was formally
applied to calculate the probability for the vacuum to remain a vacuum in a
linearly growing magnetic field for neutral fermions of spin $1/2$ with
anomalous magnetic moment. The same problem in $2+1$ dimensions was considered
in Ref. \cite{Lin99}, and in $3+1$ dimensions in Ref. \cite{LeeY06}. It is
difficult to accept the results presented in Ref. \cite{LeeY06}, which, in
particular, admit neutral-particle creation in a homogeneous magnetic field.
This means that formal calculations \emph{à la} Schwinger, without any
theoretical justification based on quantum field theory (QFT), can lead to
mistakes. The results of Ref. \cite{Lin99} seem to be reasonable, but
essentially use specific gamma matrices in $2+1$ dimensions, and cannot
provide a complete description of the effect.

It should be noted that until now a consistent description of particle
creation in the framework of QFT (due to their magnetic moments) in strong
inhomogeneous magnetic fields was unknown. To provide such a description is a
part of the present paper. In Secs. \ref{S2} and \ref{S3}, we demonstrate that
in specific cases, the problem can be technically reduced to the problem of
charged-particle creation by an electric field given by a step scalar
potential and all the information about the problem can be extracted from
exact solutions of the corresponding Dirac equation. We analyze the latter
problem once again in the framework of QFT and derive all the necessary
expressions for the probabilities of particle creation. As for the Dirac
equation, here we find a complete set of mutually commuting integrals of
motion, separate variables, and show that the energy spectrum of a neutral
fermion that interacts with an inhomogeneous magnetic field due to an
anomalous magnetic moment is real and consists of two branches separated by a
gap. In Sec. \ref{S4} we calculate all the characteristics of neutral fermion
creation from the vacuum by a linearly growing magnetic field. These results
and some of their astrophysical implications are discussed in Sec. \ref{S5}.

\section{Dirac-Pauli equation with a constant magnetic field\label{S2}}

In $3+1$ dimensions (dim.), the relativistic neutral fermions of spin $1/2$
and mass $m$ with anomalous magnetic moment $\mu$ (without an electric dipole
moment) in an external electromagnetic field $F_{\lambda\nu}$ are described by
the Dirac-Pauli equation; see Refs. \cite{Pauli41,BagGit90}. Such an equation
has the form\footnote{Here we are using the natural system of units
$\hslash=c=1$.}
\begin{align}
&  \left(  \gamma^{\lambda}\hat{p}_{\lambda}-m-\frac{1}{2}\mu\sigma
^{\lambda\nu}F_{\lambda\nu}\right)  \psi\left(  x\right)  =0\,,\nonumber\\
&  \hat{p}_{\nu}=i\partial_{\nu},\ \sigma^{\lambda\nu}=\frac{i}{2}\left[
\gamma^{\lambda},\gamma^{\nu}\right]  , \label{1}%
\end{align}
where $F_{\lambda\nu}\left(  x\right)  $ is the field tensor, $\psi\left(
x\right)  $ is a four spinor, $x=\left(  x^{0}=t,\mathbf{r}\right)
$,$\ \mathbf{r}=(x,y,z)$, and $\gamma^{\nu}=\left(  \gamma^{0}%
,\boldsymbol{\gamma}\right)  $ are Dirac matrices.

Let the external field be a constant nonuniform magnetic field $\mathbf{B}$
that is directed along the $z$ axis and depends on the coordinate $y$ only,
$\mathbf{B}\left(  y\right)  =\left(  0,0,B_{z}\left(  y\right)  \right)  $
such that the only nonzero components of the field tensor are$\ F_{21}\left(
y\right)  =-F_{12}\left(  y\right)  =B_{z}\left(  y\right)  $. In addition, we
suppose that $B_{z}\left(  y\right)  $ takes constant values as $y\rightarrow
\pm\infty$, such that
\[
\partial_{y}B_{z}\left(  y\right)  \overset{y\rightarrow\pm\infty
}{\longrightarrow}0.
\]
Moreover, we suppose that for $y<y_{L}$ (the region $S_{L}=(-\infty,y_{L}]$)
and $y>y_{R}$ (the region $S_{R}=[y_{R},\infty)$) the field $B_{z}\left(
y\right)  $ is already uniform and its values are $B_{z}\left(  y\right)
=B_{z}\left(  -\infty\right)  $ and $B_{z}\left(  y\right)  =B_{z}\left(
+\infty\right)  $, respectively. Thus, the magnetic field under consideration
is constant and uniform (or zero) at spatial infinities and, in fact,
represents either a potential barrier or step for the magnetic moment $\mu$.
With such an external field, Eq. (\ref{1}) takes the form:%
\begin{align}
&  i\partial_{0}\psi\left(  t,\mathbf{r}\right)  =\hat{H}\psi\left(
t,\mathbf{r}\right)  \,,\ \ \hat{H}=\gamma^{0}\gamma^{3}\hat{p}^{3}+\gamma
^{0}\Sigma_{z}\hat{\Pi}_{z}\,,\nonumber\\
&  \hat{\Pi}_{z}=\Sigma_{z}\boldsymbol{\gamma}\widehat{\mathbf{p}}_{\bot
}+m\Sigma_{z}-\mu B_{z}\left(  y\right)  ,\;\widehat{\mathbf{p}}_{\bot
}=\left(  \hat{p}^{1},\hat{p}^{2},0\right)  . \label{g2}%
\end{align}
In the case under consideration, the operators $\hat{p}^{0},$ $\hat{p}^{1}$,
$\hat{p}^{3}$, and $\hat{\Pi}_{z}$ are mutually commuting integrals of motion
(all these operators commute with the Hamiltonian $\hat{H}$). The integral of
motion $\hat{\Pi}_{z}$ is a generalization of the $z$ component of a spin
polarization tensor for a uniform magnetic field; see Ref. \cite{BagGit90}.

It is useful to use an additional spin operator $\hat{R}$, which is also an
integral of motion commuting with the previous ones,%
\begin{equation}
\hat{R}=\hat{H}\hat{\Pi}_{z}^{-1}\left[  1+\left(  \hat{p}^{3}\hat{\Pi}%
_{z}^{-1}\right)  ^{2}\right]  ^{-1/2}. \label{g2b}%
\end{equation}

A complete set of solutions of Eq. (\ref{g2}) can be written in the form%
\begin{equation}
\psi_{n}\left(  t,\mathbf{r}\right)  =\exp\left(  -ip_{0}t+ip_{x}%
x+ip_{z}z\right)  \psi_{n}\left(  y\right)  , \label{g3}%
\end{equation}
where $\psi_{n}\left(  y\right)  $ are eigenvectors of the equations
\begin{align}
&  s\omega\sqrt{1+\left(  p_{z}/\omega\right)  ^{2}}R\psi_{n}\left(  y\right)
=p_{0}\psi_{n}\left(  y\right)  \Longrightarrow R\psi_{n}\left(  y\right)
=s\psi_{n}\left(  y\right)  ,\;p_{0}=\omega\sqrt{1+\left(  p_{z}%
/\omega\right)  ^{2}},\nonumber\\
&  R=\left[  1+\left(  p_{z}/\omega\right)  ^{2}\right]  ^{-1/2}\left(
s\gamma^{0}\gamma^{3}p_{z}/\omega+\gamma^{0}\Sigma_{z}\right)  ,\label{g4a}\\
&  \left[  \hat{\Pi}_{z}\left(  p_{x},y\right)  -s\omega\right]  \psi
_{n}\left(  y\right)  =0,\;s=\pm1,\nonumber\\
&  \hat{\Pi}_{z}\left(  p_{x},y\right)  =\hat{\pi}_{z}-\mu B_{z}\left(
y\right)  ,\;\hat{\pi}_{z}=\Sigma_{z}\left(  \gamma^{1}p_{x}+\gamma^{2}\hat
{p}^{2}\right)  +m\Sigma_{z}, \label{g4b}%
\end{align}
and $n=(p_{x},p_{z},\omega,s)$ is the set of quantum numbers from a complete
set of numbers that will be specified below. Choosing $\psi_{n}\left(
y\right)  $ as
\[
\psi_{n}\left(  y\right)  =\frac{1}{2}\left(  1+sR\right)  \Phi\left(
y\right)  ,
\]
where $\Phi\left(  y\right)  $ is an arbitrary spinor, we obey Eq.
(\ref{g4a}).{\LARGE \ }It should be particularly emphasized that the real
continuous quantum number $\omega$ can be positive and negative and determines
the transversal part of the full energy, $\omega^{2}=p_{0}^{2}-p_{z}^{2}$,
that is, it determines the full energy of a particle moving in the $xy$ plane.
We see that the energy spectrum of the neutral fermion with anomalous momentum
is real and consists of positive and negative branches similarly to the
spectrum of the charged fermion in a time-independent electric field.

Then solutions of Eq. (\ref{g4b}) can be represented as
\begin{equation}
\psi_{n}\left(  y\right)  =\frac{1}{2}\left(  1+sR\right)  \left[  \hat{\pi
}_{z}+\mu B_{z}\left(  y\right)  +s\omega\right]  \phi_{n}\left(  y\right)  ,
\label{g5}%
\end{equation}
where the spinors $\phi_{n}\left(  y\right)  $ satisfy the following equation:%
\begin{equation}
\left\{  -\partial_{y}^{2}+m^{2}+p_{x}^{2}-\mu\gamma^{1}\partial_{y}%
B_{z}\left(  y\right)  -\left[  \omega+s\mu B_{z}\left(  y\right)  \right]
^{2}\right\}  \phi_{n}\left(  y\right)  =0. \label{g6a}%
\end{equation}

It is convenient to represent the spinor $\phi_{n}\left(  y\right)  $ in the
form%
\begin{equation}
\phi_{n}\left(  y\right)  =\varphi_{n,\chi}\left(  y\right)  \frac{1}%
{2}\left(  1+i\chi\gamma^{1}\right)  v, \label{g6c}%
\end{equation}
where it is selected that either $\chi=+1$ or $\chi=-1$, $v$ is an arbitrary
constant spinor, and the scalar functions $\varphi_{n,\chi}\left(  y\right)  $
are solutions of the equation%
\begin{equation}
\left\{  -\partial_{y}^{2}+m^{2}+p_{x}^{2}+i\chi\mu\partial_{y}B_{z}\left(
y\right)  -\left[  \omega+s\mu B_{z}\left(  y\right)  \right]  ^{2}\right\}
\varphi_{n,\chi}\left(  y\right)  =0. \label{g9}%
\end{equation}
In what follows, we suppose that $v$ is normalized as $v^{\dag}v=1$. In
addition, $vv^{\dag}$ is the identity $4\times4$ matrix, $vv^{\dag}=I$. Thus,
the spinor structure of the solutions (\ref{g5}) is defined completely. One
can easily verify that solutions (\ref{g5}) that differ by values of $\chi$
only are linearly dependent; this is an effect which the projection operator
$\left[  \ldots\right]  $ in the representation (\ref{g5}) produces. Because
of this, it is enough to work with solutions corresponding to one of two
possible values for $\chi.$ This is why the superscript $\chi$ will sometimes
disappear from solutions, but in such cases it is supposed that $\chi$ is
fixed in a certain way that is the same for all solutions under consideration.

Using the freedom inherent in the solutions of Eq. (\ref{g9}), we construct
two (in general different) sets $\left\{  _{\zeta}\psi_{n}\left(
t,\mathbf{r}\right)  \right\}  $ and $\left\{  ^{\zeta}\psi_{n}\left(
t,\mathbf{r}\right)  \right\}  $ of independent solutions, $\zeta=\pm$,
satisfying the specific boundary conditions $y\rightarrow-\infty$ or
$y\rightarrow+\infty$. The first set contains states $_{\zeta}\psi_{n}\left(
t,\mathbf{r}\right)  $ with definite real values $p^{L}$ of the $y$ component
of the momentum, such that $\zeta$ defines the sign of the momentum,%
\begin{equation}
-i\partial_{y}\ _{\zeta}\psi_{n}\left(  t,\mathbf{r}\right)  =\ p^{L}%
\ _{\zeta}\psi_{n}\left(  t,\mathbf{r}\right)  ,\ \zeta=\mathrm{sgn}%
\ p^{L},\ y\rightarrow-\infty. \label{g7a}%
\end{equation}
The second set contains states $^{\zeta}\psi_{n}\left(  t,\mathbf{r}\right)  $
with definite real values $p^{R}$ of the $y$ component of the momentum, and
again $\zeta$ defines the sign of the momentum,
\begin{equation}
-i\partial_{y}\ ^{\zeta}\psi_{n}\left(  t,\mathbf{r}\right)  =\ p^{R}%
\ ^{\zeta}\psi_{n}\left(  t,\mathbf{r}\right)  ,\ \zeta=\mathrm{sgn}%
\ p^{R},\ y\rightarrow+\infty. \label{g7b}%
\end{equation}
We are interested in the nondecaying solutions of Eq. (\ref{g9}) \textrm{\ }as
$y\rightarrow\pm\infty$. In this case\textrm{\ }both $p^{L}$ and $p^{R}$ are
real. \textrm{\ }We believe that for any given quantum numbers $n$ both sets
$\left\{  _{\zeta}\psi_{n}\left(  t,\mathbf{r}\right)  \right\}  $ and
$\left\{  ^{\zeta}\psi_{n}\left(  t,\mathbf{r}\right)  \right\}  $ represent
complete sets of nondecaying solutions. In fact this is the above-mentioned
supposition about the form of the \ field $B_{z}\left(  y\right)  $.

It should be noted that the time independence of the magnetic field under
consideration is an idealization. In fact, it is supposed that a field
inhomogeneity was switched on in a time instant $t_{in}$, which then acts as
the constant field during a large time $T$, and was switched off in a time
instant $t_{out}=t_{in}+T,$ and one can ignore the effects of its switching on
and off. This is a kind of regularization, which could---under certain
conditions---be replaced by periodic boundary conditions in $t$ . Namely, by
analogy with periodic boundary conditions in space---which are usually imposed
as the volume regularization---here we impose periodic (with the period $T$)
boundary conditions in time $t$. Thus, we consider a theory in a big
three-dimensional spacetime box that has a volume $V_{y}=TS_{xz}$%
,$\ S_{xz}=L_{x}\times L_{z}$, where $L_{x}$, $L_{z}$, and $T$ are
macroscopically large, $L_{x},L_{z}\rightarrow\infty$ and $T\rightarrow\infty$.

It is convenient to use the inner product on the time-like hyperplane
$y=\mathrm{const}$, which has the form
\begin{equation}
\left(  \psi,\psi^{\prime}\right)  _{y}=\int_{V_{y}}\psi^{\dag}\left(
t,\mathbf{r}\right)  \gamma^{0}\gamma^{2}\psi^{\prime}\left(  t,\mathbf{r}%
\right)  dtdxdz. \label{IP}%
\end{equation}
The integration in Eq. (\ref{IP}) is fulfilled in the limits from $-L_{x}/2$
to $+L_{x}/2$, $-L_{z}/2$ to $+L_{z}/2$, and from $-T/2$ to $+T/2$ in the time
$t$. It is supposed that all the functions $\psi$ are periodic under
translations from one box to another. Under these assumptions, the inner
product (\ref{IP}) does not depend on $y$. We note that the quantity
(\ref{IP}) for $\psi^{\prime}=\psi$ represents the particle current via the
hyperplane $y=\mathrm{const}$.

By using the inner product (\ref{IP}), we obtain:%
\[
\left(  \psi_{n},\psi_{k}^{\prime}\right)  _{y}=V_{y}\delta_{n,k}\psi
_{n}^{\dag}\left(  y\right)  \gamma^{0}\gamma^{2}\psi_{n}^{\prime}\left(
y\right)  .
\]
Thus, the current density in the $y$ direction in the state $\psi_{n}\left(
t,\mathbf{r}\right)  $ is
\begin{equation}
\mathcal{I}_{n}=\psi_{n}^{\dag}\left(  y\right)  \gamma^{0}\gamma^{2}\psi
_{n}\left(  y\right)  . \label{g12a}%
\end{equation}

Using the structure (\ref{g5}), we rewrite the combination $\psi_{n}^{\dag
}\left(  y\right)  \gamma^{0}\gamma^{2}\psi_{n}^{\prime}\left(  y\right)  $ as
follows
\begin{align*}
&  \varphi_{n,\chi}^{\ast}\left(  y\right)  \mathrm{tr}\left\{  \left[
-\Sigma_{z}\left(  \gamma^{1}p_{x}+i\overleftarrow{\partial}_{y}\gamma
^{2}\right)  -m\Sigma_{z}-\mu B_{z}\left(  y\right)  -s\omega\right]
\gamma^{0}\gamma^{2}\frac{1}{2}\left[  1+sR\right]  \right. \\
&  \left.  \left[  -\Sigma_{z}\left(  \gamma^{1}p_{x}%
-i\overrightarrow{\partial}_{y}\gamma^{2}\right)  -m\Sigma_{z}-\mu
B_{z}\left(  y\right)  -s\omega\right]  \frac{1}{2}\left(  1+\chi i\gamma
^{1}\right)  \right\}  \varphi_{n,\chi}^{\prime}\left(  y\right)  ,
\end{align*}
where $\mathrm{tr}\left\{  ...\right\}  $ is the trace in the space of
$4\times4$ matrices. Calculating this trace, we obtain%
\begin{align}
\psi_{n}^{\dag}\left(  y\right)  \gamma^{0}\gamma^{2}\psi_{n}^{\prime}\left(
y\right)   &  =\left(  1+\left(  p_{z}/\omega\right)  ^{2}\right)
^{-1/2}\varphi_{n,\chi}^{\ast}\left(  y\right)  \left(
i\overleftarrow{\partial}_{y}-i\overrightarrow{\partial}_{y}\right)
\nonumber\\
&  \times\left(  \omega+s\mu B_{z}\left(  y\right)  +s\chi
i\overrightarrow{\partial}_{y}\right)  \varphi_{n,\chi}^{\prime}\left(
y\right)  . \label{g12}%
\end{align}

As was already mentioned, we supposed that $B_{z}\left(  y\right)  $ tends to
some constant values as $y\rightarrow\pm\infty$. Let us suppose for sake of
definiteness that the derivative $\partial_{y}B_{z}(y)$ has a definite sign,
let us say $\partial_{y}B_{z}(y)\geq0$, $\forall y$, and let $\mu<0$. Note
that there are no bound states in this case. To simplify the consideration, we
also suppose that
\[
U=U_{R}-U_{L}>0,\ U_{L}=-\mu B_{z}\left(  -\infty\right)  <0,\ U_{R}=-\mu
B_{z}\left(  +\infty\right)  >0.
\]
For asymptotic (as $\left\vert y\right\vert \rightarrow\infty$) states with
real values $p^{L}$ and/or $p^{R}$, we have
\begin{equation}
_{\zeta}\varphi_{n,\chi}\left(  y\right)  =\;_{\zeta}\mathcal{N}\exp\left(
ip^{L}y\right)  ,\;\;^{\zeta}\varphi_{n}^{\left(  \chi\right)  }\left(
x\right)  =\;^{\zeta}\mathcal{N}\exp\left(  ip^{R}y\right)  , \label{e8a}%
\end{equation}
respectively, where $_{\zeta}\mathcal{N}$ and $^{\zeta}\mathcal{N}$ are
normalization factors. We introduce the notation%
\begin{align}
E_{s}\left(  L/R\right)   &  =\pi_{s}\left(  L/R\right)  \sqrt{1+\left[
p_{z}/\pi_{s}\left(  L/R\right)  \right]  ^{2}},\nonumber\\
\pi_{s}\left(  L/R\right)   &  =\omega-sU_{L/R},\;\pi_{x}=\sqrt{p_{x}%
^{2}+m^{2}} \label{e9}%
\end{align}
and in their terms we stress the existence of the following relations
\begin{align}
&  \pi_{s}\left(  L\right)  =\pi_{s}\left(  R\right)  +sU,\label{ca}\\
&  \left(  p^{L}\right)  ^{2}=\left[  E_{s}\left(  L\right)  \right]  ^{2}%
-\pi_{x}^{2}-p_{z}^{2},\ \left(  p^{R}\right)  ^{2}=\left[  E_{s}\left(
R\right)  \right]  ^{2}-\pi_{x}^{2}-p_{z}^{2}, \label{ca0}%
\end{align}
where Eq. (\ref{ca0}) holds due to Eq. (\ref{g9}). We see that $\left\vert
E_{s}\left(  L\right)  \right\vert $ and $\left\vert E_{s}\left(  R\right)
\right\vert $ are the asymptotic values of the kinetic energy, while
$\left\vert \pi_{s}\left(  L\right)  \right\vert $ and $\left\vert \pi
_{s}\left(  R\right)  \right\vert $ are the asymptotic values of its
transversal part, respectively.

Note that the case of the uniform magnetic field is realized when
$U_{R}\rightarrow U_{L}=-\mu B_{z}$; then, asymptotic regions coincide and
coincide with the whole space as well, $\pi_{s}\left(  L\right)  =\pi
_{s}\left(  R\right)  =\omega+s\mu B_{z}$, and $p^{L}=p^{R}=p_{y}$. It follows
from Eqs. (\ref{ca0}) that%
\[
p_{x}^{2}+p_{y}^{2}+m^{2}=\left(  \omega+s\mu B_{z}\right)  ^{2}%
\Longrightarrow\omega+s\mu B_{z}=\pm\sqrt{p_{x}^{2}+p_{y}^{2}+m^{2}}%
\]
and we see that $\left\vert \omega+s\mu B_{z}\right\vert $ is the transversal
part of the kinetic energy. Thus, using standard second quantization, we can
construct the Fock space of fermions with conserved spin polarization $s$,
where $\omega+s\mu B_{z}\geq m$\ for\ particles, and\ $\omega+s\mu B_{z}%
\leq-m\;$for\ antiparticles. One can see that in contrast to the statement of
Ref. \cite{LeeY06}---which is the result of an improper treatment of naive
spectra that, in fact, are valued for the case of a weak field; see Ref.
\cite{BagGit90}---the energy spectrum of neutral fermions interacting with
uniform magnetic field due to an anomalous magnetic moments is real and a
level crossing and vacuum instability is absent. In fact, this Fock space is
equivalent to the Fock space of free particles.

Then, using the asymptotic conditions (\ref{g7a}) and (\ref{g7b}), and the
result (\ref{g12}), we can subject the introduced sets $\left\{  _{\zeta}%
\psi_{n}\left(  t,\mathbf{r}\right)  \right\}  $ and $\left\{  ^{\zeta}%
\psi_{n}\left(  t,\mathbf{r}\right)  \right\}  $ to the following
orthonormality conditions%
\begin{equation}
\left(  _{\zeta}\psi_{n},_{\zeta^{\prime}}\psi_{n^{\prime}}\right)  _{x}%
=\zeta\eta_{L}\delta_{\zeta,\zeta^{\prime}}\delta_{n,n^{\prime}};\;\;\left(
^{\zeta}\psi_{n},^{\zeta^{\prime}}\psi_{n^{\prime}}\right)  _{x}=\zeta\eta
_{R}\delta_{\zeta,\zeta^{\prime}}\delta_{n,n^{\prime}}, \label{g13}%
\end{equation}
where%
\[
\eta_{L}=\mathrm{sgn}\pi_{s}\left(  L\right)  ,\ \ \eta_{R}=\mathrm{sgn}%
\pi_{s}\left(  R\right)  .
\]
In deriving Eq. (\ref{g13}), it was taken into account that for asymptotic (as
$\left\vert y\right\vert \rightarrow\infty$) states with real values $p^{L}$
and $p^{R}$, the relations%
\[
\left\vert \pi_{s}\left(  L\right)  \right\vert >\left\vert p^{L}\right\vert
,\ \left\vert \pi_{s}\left(  R\right)  \right\vert >\left\vert p^{R}%
\right\vert
\]
hold due to Eq. (\ref{ca0}), respectively. This is why the sign of the
quantity (\ref{g12}) with the operator $\left[  \pi_{s}\left(  L/R\right)
+s\chi i\overrightarrow{\partial}_{y}\right]  $ is due to the sign of the
$\pi_{s}\left(  L/R\right)  $. The normalization factors in Eq. (\ref{e8a})
are as follows%
\begin{align}
_{\zeta}\mathcal{N}  &  \mathcal{=}\;_{\zeta}CY,\;\;^{\zeta}\mathcal{N=}%
\;^{\zeta}CY,\;\;Y=\left(  1+\left(  p_{z}/\omega\right)  ^{2}\right)
^{1/4}V_{y}^{-1/2},\nonumber\\
\;_{\zeta}C  &  =\left[  2\left\vert p^{L}\right\vert \left\vert \pi
_{s}\left(  L\right)  -s\chi p^{L}\right\vert \right]  ^{-1/2},\;\;\;^{\zeta
}C=\left[  2\left\vert p^{R}\right\vert \left\vert \pi_{s}\left(  R\right)
-s\chi p^{R}\right\vert \right]  ^{-1/2}. \label{e8b}%
\end{align}
In the limit of infinite volume of the normalization (continuous momenta
$p_{0}$, $p_{x}$, and $p_{z}$) one has to substitute $\delta_{n,n^{\prime}}$
into the normalization conditions (\ref{g13}) by $\delta_{s,s^{\prime}}%
\delta\left(  p_{0}-p_{0}^{\prime}\right)  \delta\left(  p_{x}-p_{x}^{\prime
}\right)  \delta\left(  p_{z}-p_{z}^{\prime}\right)  .$ In this case,
$V_{y}^{-1/2}\rightarrow\left(  2\pi\right)  ^{-3/2}$ in Eqs. (\ref{e8b}).

It is supposed that for any given quantum numbers $n$, both sets $\left\{
_{\zeta}\psi_{n}\left(  t,\mathbf{r}\right)  \right\}  $ and $\left\{
^{\zeta}\psi_{n}\left(  t,\mathbf{r}\right)  \right\}  $ represent complete
sets of nondecaying solutions of Eq. (\ref{g2}). Then their mutual
decompositions have the form%
\begin{align}
\eta_{L}\;^{\;\zeta}\psi_{n}\left(  t,\mathbf{r}\right)   &  =\;_{+}\psi
_{n}\left(  t,\mathbf{r}\right)  g\left(  _{+}\left\vert ^{\zeta}\right.
\right)  -\;_{-}\psi_{n}\left(  t,\mathbf{r}\right)  g\left(  _{-}\left\vert
^{\zeta}\right.  \right)  ;\nonumber\\
\eta_{R}\;_{\zeta}\psi_{n}\left(  t,\mathbf{r}\right)   &  =\;^{\;+}\psi
_{n}\left(  t,\mathbf{r}\right)  g\left(  ^{+}\left\vert _{\zeta}\right.
\right)  \;-\;^{\;-}\psi_{n}\left(  t,\mathbf{r}\right)  g\left(
^{-}\left\vert _{\zeta}\right.  \right)  \;, \label{rel1}%
\end{align}
where the decomposition coefficients $g$ are defined by the relations:
\begin{equation}
\left(  _{\zeta}\psi_{n},\;^{\zeta^{\prime}}\psi_{n^{\prime}}\right)
_{y}=\delta_{nn^{\prime}}g\left(  _{\zeta}\left\vert ^{\zeta^{\prime}}\right.
\right)  ,\;g\left(  ^{\zeta^{\prime}}\left\vert _{\zeta}\right.  \right)
=g\left(  _{\zeta}\left\vert ^{\zeta^{\prime}}\right.  \right)  ^{\ast}.
\label{INy}%
\end{equation}
Using the orthonormality conditions (\ref{g13}), we derive the following
relations for the decomposition coefficients:%
\begin{align}
&  g\left(  ^{\zeta^{\prime}}\left\vert _{+}\right.  \right)  g\left(
_{+}\left\vert ^{\zeta}\right.  \right)  -g\left(  ^{\zeta^{\prime}}\left\vert
_{-}\right.  \right)  g\left(  _{-}\left\vert ^{\zeta}\right.  \right)
=\zeta\eta_{L}\eta_{R}\delta_{\zeta,\zeta^{\prime}};\nonumber\\
&  g\left(  _{\zeta^{\prime}}\left\vert ^{+}\right.  \right)  g\left(
^{+}\left\vert _{\zeta}\right.  \right)  -g\left(  _{\zeta^{\prime}}\left\vert
^{-}\right.  \right)  g\left(  ^{-}\left\vert _{\zeta}\right.  \right)
=\zeta\eta_{L}\eta_{R}\delta_{\zeta,\zeta^{\prime}}. \label{UR}%
\end{align}
In particular, these relations imply that%
\begin{equation}
\left\vert g\left(  _{-}\left\vert ^{+}\right.  \right)  \right\vert
^{2}=\left\vert g\left(  _{+}\left\vert ^{-}\right.  \right)  \right\vert
^{2},\;\left\vert g\left(  _{+}\left\vert ^{+}\right.  \right)  \right\vert
^{2}=\left\vert g\left(  _{-}\left\vert ^{-}\right.  \right)  \right\vert
^{2},\;\frac{g\left(  _{+}\left\vert ^{-}\right.  \right)  }{g\left(
_{-}\left\vert ^{-}\right.  \right)  }=\frac{g\left(  ^{+}\left\vert
_{-}\right.  \right)  }{g\left(  ^{+}\left\vert _{+}\right.  \right)  }.
\label{UR2}%
\end{equation}
Thus, one can see that all these coefficients can be expressed via only two of
them, e.g. via $g\left(  _{+}\left\vert ^{+}\right.  \right)  $ and $g\left(
_{+}\left\vert ^{-}\right.  \right)  $. However, even these coefficients are
not completely independent, they are related as follows:
\begin{equation}
\left\vert g\left(  _{+}\left\vert ^{-}\right.  \right)  \right\vert
^{2}-\left\vert g\left(  _{+}\left\vert ^{+}\right.  \right)  \right\vert
^{2}=-\eta_{L}\eta_{R}. \label{UR1}%
\end{equation}

\section{\textbf{Creation of neutral fermions}\label{S3}}

It is useful to make a preliminary qualitative analysis of the behavior of
particles and antiparticles in the fields under consideration. It should be
noted that here there exist two principally different cases, the first one
corresponds to $U<2m$, whereas the second one (we call it the creation case,
or C-case) corresponds to $U>2m$. In the first case, there exist only a
scattering of neutral fermions by the magnetic field without additional
particle creation from the vacuum. This case can be treated in the framework
of one-particle relativistic quantum mechanics. The quantum number $s$ gives
the spin polarization for both particles and antiparticles. Choosing the
magnetic moment of the particle as $\mu$, we have the magnetic moment of the
antiparticle as $-\mu$. Note that we fix $\mu=-\left\vert \mu\right\vert $.
Then, according to the standard particle-antiparticle identification of wave
functions, the asymptotic kinetic energy (at $y\rightarrow\pm\infty$) of the
particle moving in the $xy$ plane is $\pi_{s}\left(  L/R\right)  >0,$ while it
is $-\pi_{s}\left(  L/R\right)  >0$ for the antiparticle. One can see from Eq.
(\ref{e9}) that the particle potential energy $s\left\vert \mu\right\vert
B_{z}\left(  y\right)  $ decreases along the $y$ axis for $s=-1$ and increases
for $s=+1$. At the same time, the antiparticle potential energy $-s\left\vert
\mu\right\vert B_{z}\left(  y\right)  $ increases along the $y$ axis for
$s=-1$ and decreases for $s=+1$. This means that the field $B_{z}\left(
y\right)  $ accelerates particles with $s=-1$ and antiparticles with $s=+1$
along the $y$ axis. Respectively, antiparticles with $s=-1$ and particles with
$s=+1$ are accelerated by the field in the opposite direction. The same
observation holds in the case $U>2m.$

We note that real particles are described by some wave packets localized in
the spacetime , such that we have to study the motion of such packets in the
external field (obviously, it is enough to speak about a localization in the
$y$ direction). Let us denote by $S_{\mathrm{int}}$ the region, where the
magnetic field is inhomogeneous. In the region $S_{L},$ situated \ to the left
of $S_{\mathrm{int}}$ and in the region $S_{R}$ to the right of
$S_{\mathrm{int}}$, the magnetic field is homogeneous. For big enough
differences $U$ between the initial and final potential energies, particles
and antiparticles with any initial kinetic momenta along the $y$ axis get
final kinetic momenta that is always in the same direction as their
acceleration by the magnetic field. This is what we have in the case
$U>2\sqrt{p_{x}^{2}+p_{z}^{2}+m^{2}}$ for all partial waves with given $p_{x}$
and $p_{z}$ of a wave packet. Because particles and their antiparticles with a
given $s$ have opposite directions of acceleration, there exists a state
polarization out of the region $S_{\mathrm{int}}$. The final particles with
$s=+1$ and antiparticles with $s=-1$ are situated in the region $S_{L}$, and
final antiparticles with $s=+1$ and particles with $s=-1$ are situated in the
region $S_{R}$.

From the physical point of view, there is a similarity between the two
cases---one where neutral fermions with an anomalous magnetic moment are
placed in an inhomogeneous magnetic field $B_{z}\left(  y\right)  $ with
$\partial_{y}B_{z}(y)>0$, and another where charged fermions are placed in a
constant electric field directed along $y$ and given by a scalar potential
$A_{0}\left(  y\right)  $. In both cases external fields produce a work which
implies an acceleration of the corresponding particles in the $y$ direction.
From the QFT point of view if such a work is greater than $2m$ (C-case),
particle creation from the vacuum is possible. In fact, this analogy allows in
both cases formally to use the same techniques of calculation. It turns out
that the problem of neutral fermion creation in strong inhomogeneous magnetic
field can be technically reduced to the problem of charged-particle creation
by an electric potential step. Some heuristic exact calculations of the
particle creation by potential steps in the framework of the relativistic
quantum mechanics were presented by Nikishov \cite{Nik-barrier-70,Nikis79},
further developed in Ref. \cite{HansRavn81}, and used in numerous works in the
framework of semiclassical considerations; for a review see Refs.
\cite{DomCal99,RufVSh10}\textrm{. }

In such a way it seems that we could use the known results to find the mean
number of neutral particle-antiparticle pairs created. However, a closer
consideration shows that the\textrm{\ }particle-antiparticle and causal
identification of wave functions $^{\zeta}\psi_{n}\left(  t,\mathbf{r}\right)
$ and $_{\zeta}\psi_{n}\left(  t,\mathbf{r}\right)  $ given by Nikishov
\cite{Nik-barrier-70,Nikis79} does not coincide with that given by Hansen and
Ravndal \cite{HansRavn81} for the C-case; see the discussion in Ref.
\cite{Nik04}\textrm{. }Within the WKB approximation this difficulty can be
bypassed, but the question remains. Trying to resolve this contradiction, we
have realized that at that time no justification for quantum mechanical
calculations from the QFT point of view were elaborated. Such a justification
can be obtained in the framework of a strict QFT formulation of particle
creation by potential steps; see our forthcoming work \cite{GavGit12}. Here
for our specific purposes it is enough to use the solution presented above,
taking into account some necessary physical considerations.

In the C-case, there exists a range $2\sqrt{\pi_{x}^{2}+p_{z}^{2}}<U$ of the
momenta $p_{x}$ and $p_{z}$ of the fermions, such that particle creation is
possible. This case is described by the wave functions (\ref{g5}) with quantum
numbers from the range $\Omega$, where $\omega$, $p_{x}$, and $p_{z}$ are
restricted by the inequalities%
\begin{equation}
\Omega:\ s\pi_{s}\left(  L\right)  \geq\pi_{x},\;s\pi_{s}\left(  R\right)
\leq-\pi_{x},\;2\sqrt{\pi_{x}^{2}+p_{z}^{2}}<U. \label{g14}%
\end{equation}
If we treat this case using the identification of a wave function by an
analogy with one-particle scattering theory, there appears an analog of the
Klein paradox for charged relativistic particles in an electric field
\cite{Klein29}. This is an indication that one has to use an appropriate
many-particle description given by QFT to treat the problem correctly.

In the first stage of the canonical quantization of the field $\psi\left(
t,\mathbf{r}\right)  $ one establishes that the corresponding quantum field is
the Heisenberg field operator $\Psi\left(  t,\mathbf{r}\right)  $ that
satisfies the equal-time anticommutation relations:%
\begin{equation}
\left[  \Psi\left(  t,\mathbf{r}\right)  ,\Psi\left(  t,\mathbf{r}^{\prime
}\right)  \right]  _{+}=\left[  \Psi\left(  t,\mathbf{r}\right)  ^{\dagger
},\Psi\left(  t,\mathbf{r}^{\prime}\right)  ^{\dagger}\right]  _{+}%
=0,\ \left[  \Psi\left(  t,\mathbf{r}\right)  ,\Psi\left(  t,\mathbf{r}%
^{\prime}\right)  ^{\dagger}\right]  _{+}=\delta\left(  \mathbf{r-r}^{\prime
}\right)  \label{3.3}%
\end{equation}
and the Dirac-Pauli equation (\ref{g2}). The formal expressions for the
Hamiltonian $\widehat{\mathcal{H}}$ of the quantized fermion field and the
corresponding magnetic momentum operator $\widehat{\mathcal{M}}$ can be easily
constructed,%
\begin{equation}
\widehat{\mathcal{H}}=\int\Psi\left(  t,\mathbf{r}\right)  ^{\dagger}\hat
{H}\Psi\left(  t,\mathbf{r}\right)  d\mathbf{r,\ }\widehat{\mathcal{M}}%
=\frac{\mu}{2}\int\left[  \Psi\left(  t,\mathbf{r}\right)  ^{\dagger}%
,\Psi\left(  t,\mathbf{r}\right)  \right]  _{-}d\mathbf{r}. \label{3.4}%
\end{equation}

To perform quantization in terms of particles and antiparticles, we define the
inner product
\begin{equation}
\left(  \psi,\psi^{\prime}\right)  _{t}=\int_{t}\psi^{\dag}\left(
t,\mathbf{r}\right)  \psi^{\prime}\left(  t,\mathbf{r}\right)  d\mathbf{r}
\label{t3}%
\end{equation}
between two solutions of the the Dirac-Pauli equation on a $t=\mathrm{const}$
hyperplane. This inner product does not depend on the choice of such a
hyperplane if the spinors $\psi\left(  t,\mathbf{r}\right)  $ obey certain
boundary conditions that allow one to integrate by parts in Eq. (\ref{t3})
neglecting boundary terms. Since physical states are wave packets that vanish
on the remote boundaries, the above assumption holds true and the inner
product (\ref{t3}) is time independent for such states. Considering plane
waves instead of natural wave packets, one has to impose corresponding
periodic boundary conditions on the corresponding wave functions and the
external field to keep the inner product (\ref{t3}) time independent. However,
in the case under consideration the external field with different asymptotics
at $y\rightarrow\pm\infty$ cannot be adapted to any periodic boundary
conditions in the $y$ direction without changing its physical content. To
provide time independence of the inner product, one has to redefine\textbf{\ }%
the inner product itself. This modification is applied to the integration over
$y$ in the expression (\ref{t3}) and is described below.

Let $\psi_{n}\left(  t,\mathbf{r}\right)  $ and $\psi_{n^{\prime}}^{\prime
}\left(  t,\mathbf{r}\right)  $ be wave functions (\ref{g5}) and the integral
over the variable $y$\ in the infinite limits be regularized by large positive
numbers $L_{1}$ and $L_{2}$. Integrating over the variables $x$, $z$, and
using representation (\ref{g6c}), we obtain
\begin{align}
&  \left(  \psi_{n},\psi_{n^{\prime}}^{\prime}\right)  _{t}=\delta
_{n,n^{\prime}}S_{xz}\mathcal{R},\;\;\mathcal{R}=\int_{-L_{1}}^{L_{2}%
}Qdy,\nonumber\\
&  Q=\left(  \varphi_{n,\chi}\left(  y\right)  \right)  ^{\ast}\left[  \pi
_{x}^{2}+\left(  \omega+s\mu B_{z}\left(  y\right)  +s\chi i\partial
_{y}\right)  ^{2}\right]  \varphi_{n,\chi}^{\prime}\left(  y\right)  ,
\label{i1}%
\end{align}
where the orthogonality for $n\neq n^{\prime}$ follows as $L_{1},$
$L_{2}\rightarrow\infty$.

We represent the regularized integral $\mathcal{R}$ as%
\begin{equation}
\mathcal{R=}\int_{-L_{1}}^{y_{L}}Qdy+\int_{y_{L}}^{y_{R}}Qdy+\int_{y_{R}%
}^{L_{2}}Qdy, \label{i2}%
\end{equation}
where only the second term---the integral over the region $S_{\mathrm{int}}%
$---depends on the derivative $\partial_{y}B_{z}(y)$. The smoothness of the
$\partial_{y}B_{z}(y)$ allows us to believe that this integral is finite as
$L_{1},$ $L_{2}\rightarrow\infty$. The first and the third terms are
calculated as integrals over the regions where $\partial_{y}B_{z}(y)=0$. Then
their values are determined by the asymptotics (\ref{e8a}) in the following
form%
\begin{align}
\mathcal{R}_{L}  &  =\int_{-L_{1}}^{y_{L}}Q_{L}dy,\;\;\mathcal{R}_{R}%
=\int_{y_{R}}^{L_{2}}Q_{R}dy,\nonumber\\
Q_{L/R}  &  =\left(  \varphi_{n,\chi}\left(  y\right)  \right)  ^{\ast}\left[
\pi_{x}^{2}+\left(  \pi_{s}\left(  L/R\right)  +s\chi i\partial_{y}\right)
^{2}\right]  \varphi_{n,\chi}^{\prime}\left(  y\right)  . \label{i3}%
\end{align}
$Q_{L}$ and $Q_{R}$ are constant then $\mathcal{R}_{L}\sim L_{1}$ and
$\mathcal{R}_{R}\sim L_{2}$. We see that only $\mathcal{R}_{L}$ and
$\mathcal{R}_{R}$ make a contribution to $\mathcal{R}$ in Eq. (\ref{i2}) as
$L_{1},$ $L_{2}\rightarrow\infty$,
\[
\mathcal{R}\underset{L_{1},L_{2}\rightarrow\infty}{\longrightarrow}%
\mathcal{R}_{L}+\mathcal{R}_{R}\ .
\]

There exist two independent solutions with a given quantum number $n$ from the
range $\Omega$. In spite of the fact that these solutions are obtained in the
constant external field we believe that they represent asymptotic forms of
some unknown solutions of the Dirac-Pauli equation with an external field
$\partial_{y}B_{z}(t,y)$ that is switched on and off at $t\rightarrow\pm
\infty$ and the effects of the switching from on to off are negligible. Since
the inner product (\ref{t3}) does not depend on $t$ for such solutions, we
believe that orthogonal pairs of solutions that describe alternative
particle/antiparticle states at the initial and the final time instants remain
orthogonal at an arbitrary instant of time. Therefore we have to find out
which solutions among those we have introduced before are such orthogonal
pairs. Taking into account the relations (\ref{rel1}), one can show that%
\begin{equation}
\left(  _{\zeta}\psi_{n},^{\zeta}\psi_{n}\right)  _{t}=0,\;\;n\in
\Omega,\label{i9}%
\end{equation}
if we assume that $L_{1}$ and $L_{2}$ satisfy the relation%

\begin{equation}
L_{1}\left\vert \frac{\pi_{s}\left(  L\right)  }{p^{L}}\right\vert
-L_{2}\left\vert \frac{\pi_{s}\left(  R\right)  }{p^{R}}\right\vert =O\left(
1\right)  . \label{i8}%
\end{equation}
Condition (\ref{i8}) guarantees that the wave functions $_{\zeta}\psi
_{n}\left(  t,\mathbf{r}\right)  $ and $^{\zeta}\psi_{n}\left(  t,\mathbf{r}%
\right)  $ for $n\in\Omega$ correspond to alternative physical
states.\textrm{\ }Note that condition (\ref{i8}) is unique to guarantee that
all the wave functions with any $n$ of the complete set corresponding to
alternative physical states are orthogonal with respect to the inner product
(\ref{t3}); for details see our forthcoming work \cite{GavGit12}. In fact such
a condition has to be considered as a part of the definition of the inner
product (\ref{t3}).

Consider the quantities $\mathcal{R}_{L/R}$ (\ref{i3}) defined by the
functions $_{\zeta}\varphi_{n}\left(  x\right)  $ and $^{\zeta}\varphi
_{n}\left(  x\right)  $ with quantum numbers $n$ from the range $\Omega.$ In
this case we attribute the corresponding index $\zeta$ to these quantities as
follows: $\mathcal{R}_{L/R}\rightarrow\ _{\zeta}\mathcal{R}_{L/R}$ or
$\mathcal{R}_{L/R}\rightarrow\ ^{\zeta}\mathcal{R}_{L/R}.$ Using Eqs.
(\ref{e8a}) and (\ref{e8b}) and retaining only leading terms in the limit
$L_{1},L_{2}\rightarrow\infty$, we obtain%
\begin{equation}
_{\zeta}\mathcal{R}_{L}=Y^{2}L_{1}\left\vert \frac{\pi_{s}\left(  L\right)
}{p^{L}}\right\vert ,\;\;^{\zeta}\mathcal{R}_{R}=Y^{2}L_{2}\left\vert
\frac{\pi_{s}\left(  R\right)  }{p^{R}}\right\vert . \label{i10}%
\end{equation}
To calculate the quantities $_{\zeta}\mathcal{R}_{R}$ and $^{\zeta}%
\mathcal{R}_{L}$, we use the relations (\ref{rel1}). Again retaining only
leading terms in the limit $L_{1},L_{2}\rightarrow\infty$ (neglecting in
particular oscillating terms) and taking into account Eqs. (\ref{e8a}) and
(\ref{e8b}), we find
\begin{align}
_{\zeta}\mathcal{R}_{R}  &  =Y^{2}L_{2}\left\vert \frac{\pi_{s}\left(
R\right)  }{p^{R}}\right\vert \left[  \left\vert g\left(  _{\zeta}\left\vert
^{+}\right.  \right)  \right\vert ^{2}+\left\vert g\left(  _{\zeta}\left\vert
^{-}\right.  \right)  \right\vert ^{2}\right]  ,\nonumber\\
^{\zeta}\mathcal{R}_{L}  &  =Y^{2}L_{1}\left\vert \frac{\pi_{s}\left(
L\right)  }{p^{L}}\right\vert \left[  \left\vert g\left(  _{+}\left\vert
^{\zeta}\right.  \right)  \right\vert ^{2}+\left\vert g\left(  _{-}\left\vert
^{\zeta}\right.  \right)  \right\vert ^{2}\right]  . \label{i11}%
\end{align}
Note that $^{\zeta}\mathcal{R}_{L}>\;^{\zeta}\mathcal{R}_{R}$ and $_{\zeta
}\mathcal{R}_{R}>\;_{\zeta}\mathcal{R}_{L}$ due to $\left\vert g\left(
_{+}\left\vert ^{-}\right.  \right)  \right\vert ^{2}>1$. Taking the unitarity
relations (\ref{UR1}) and the condition (\ref{i8}) into account, we obtain the
following orthonormality relations
\begin{align}
&  \left(  _{\zeta}\psi_{n},_{\zeta}\psi_{n^{\prime}}\right)  _{t}%
=\delta_{n,n^{\prime}}C_{t},\;\;\left(  ^{\zeta}\psi_{n},^{\zeta}%
\psi_{n^{\prime}}\right)  _{t}=\delta_{n,n^{\prime}}C_{t},\nonumber\\
&  C_{t}=2\frac{L_{2}}{T}\left\vert \frac{\pi_{s}\left(  R\right)  }{p^{R}%
}\right\vert \left\vert g\left(  _{+}\left\vert ^{-}\right.  \right)
\right\vert ^{2}. \label{i12}%
\end{align}

One can see that the following symmetry occurs: particles with opposite values
of $s$ have opposite accelerations; the same is valid for antiparticles. This
is why the cases $s=+1$ and $s=-1$ differ only by opposite directions of all
the motions, and respectively by the opposite dispositions of all the
asymptotic ranges. The probabilities of all the processes are equal in both
the cases. This is why it is enough to consider only one case, let us say
$s=+1$.

It is supposed that we know the complete set of the solutions of the
Dirac-Pauli equation, parametrized by a set of quantum numbers $n$, on the
hyperplane $t=\mathrm{const}$. Then we can decompose the quantum Heisenberg
field operator $\Psi\left(  t,\mathbf{r}\right)  $ and its Hermitian conjugate
$\Psi^{\dag}\left(  t,\mathbf{r}\right)  $ in this complete set using the
inner product (\ref{t3}). Assuming that both sets $\left\{  _{+}\psi
_{n}\left(  t,\mathbf{r}\right)  ,^{+}\psi_{n}\left(  t,\mathbf{r}\right)
\right\}  $ and $\left\{  _{-}\psi_{n}\left(  t,\mathbf{r}\right)  ,^{-}%
\psi_{n}\left(  t,\mathbf{r}\right)  \right\}  $ represent the complete set of
nondecaying solutions in the range $\Omega$, we introduce the notation
$\Psi_{n}\left(  t,\mathbf{r}\right)  $ for the component of the quantum field
operator that can be expanded via either $_{+}\psi_{n}\left(  t,\mathbf{r}%
\right)  ,^{+}\psi_{n}\left(  t,\mathbf{r}\right)  $ or $_{-}\psi_{n}\left(
t,\mathbf{r}\right)  ,^{-}\psi_{n}\left(  t,\mathbf{r}\right)  $. Operator
coefficients in such decompositions do not depend on spacetime coordinates
because both quantum field operators and classical solutions obey the same
Pauli-Dirac equation. For example, for $s=+1$, we can decompose the $\Psi
_{n}\left(  t,\mathbf{r}\right)  $ and $\Psi_{n}^{\dagger}\left(
t,\mathbf{r}\right)  $ as follows
\begin{align}
&  \Psi_{n}\left(  t,\mathbf{r}\right)  =C_{t}^{-1/2}\left[  a_{n}\left(
\mathrm{out}\right)  \;^{+}\psi_{n}\left(  t,\mathbf{r}\right)  +b_{n}^{\dag
}\left(  \mathrm{out}\right)  \;_{+}\psi_{n}\left(  t,\mathbf{r}\right)
\right]  ,\nonumber\\
&  \Psi_{n}^{\dag}\left(  t,\mathbf{r}\right)  =C_{t}^{-1/2}\left[
a_{n}^{\dagger}\left(  \mathrm{out}\right)  \;^{+}\psi_{n}^{\dagger}\left(
t,\mathbf{r}\right)  +b_{n}\left(  \mathrm{out}\right)  \;_{+}\psi
_{n}^{\dagger}\left(  t,\mathbf{r}\right)  \right]  ; \label{c5}%
\end{align}
and%
\begin{align}
&  \;\Psi_{n}\left(  t,\mathbf{r}\right)  =C_{t}^{-1/2}\left[  a_{n}\left(
\mathrm{in}\right)  \;^{-}\psi_{n}\left(  t,\mathbf{r}\right)  +b_{n}^{\dag
}\left(  \mathrm{in}\right)  \;_{-}\psi_{n}\left(  t,\mathbf{r}\right)
\right]  ,\nonumber\\
&  \Psi_{n}^{\dag}\left(  t,\mathbf{r}\right)  =C_{t}^{-1/2}\left[
a_{n}^{\dagger}\left(  \mathrm{in}\right)  \;^{-}\psi_{n}^{\dagger}\left(
t,\mathbf{r}\right)  +b_{n}\left(  \mathrm{in}\right)  \;_{-}\psi_{n}%
^{\dagger}\left(  t,\mathbf{r}\right)  \right]  . \label{c6}%
\end{align}
In what follows, we interpret all $a$ and $b\ $as annihilation and all
$a^{\dag}$ and $b^{\dag}$ as creation operators; all $a$ and $a^{\dag}$ as
describing particles and $b$ and $b^{\dag}$ as describing antiparticles, and
all the operators labeled by the argument "$\mathrm{in}$" are $\mathrm{in}%
$-operators, whereas all the operators labeled by the argument "$\mathrm{out}%
$" are $\mathrm{out}$-operators. \ It can be shown that these creation and
annihilation operators obey canonical anticommutation relations,%
\begin{align}
\lbrack a_{n}(\mathrm{in}),a_{k}^{\dagger}(\mathrm{in})]_{+}  &
=[a_{n}(\mathrm{out}),a_{k}^{\dagger}(\mathrm{out})]_{+}=[b_{n}(\mathrm{in}%
),b_{k}^{\dagger}(\mathrm{in})]_{+}=[b_{n}(\mathrm{out}),b_{k}^{\dagger
}(\mathrm{out})]_{+}=\delta_{n,k},\nonumber\\
\lbrack a_{n}(\mathrm{out}),a_{k}(\mathrm{out})]_{+}  &  =[b_{n}%
(\mathrm{out}),b_{k}(\mathrm{out})]_{+}=[a_{n}(\mathrm{out}),b_{k}%
(\mathrm{out})]_{+}=[a_{n}(\mathrm{out}),b_{k}^{\dagger}(\mathrm{out}%
)]_{+}=0,\nonumber\\
\lbrack a_{n}(\mathrm{in}),a_{k}(\mathrm{in})]_{+}  &  =[b_{n}(\mathrm{in}%
),b_{k}(\mathrm{in})]_{+}=[a_{n}(\mathrm{in}),b_{k}(\mathrm{in})]_{+}%
=[a_{n}(\mathrm{in}),b_{k}^{\dagger}(\mathrm{in})]_{+}=0,\ \label{c8}%
\end{align}
due to relation (\ref{3.3}). In such an interpretation, the $\mathrm{in}%
$-vacuum $\left\vert 0,\mathrm{in}\right\rangle $ and $\mathrm{out}$-vacuum
$\left\vert 0,\mathrm{out}\right\rangle $ are defined by the conditions,%
\begin{align}
&  a_{n}\left(  \mathrm{in}\right)  \left\vert 0,\mathrm{in}\right\rangle
=b_{n}\left(  \mathrm{in}\right)  \left\vert 0,\mathrm{in}\right\rangle
=0,\;\forall n;\nonumber\\
&  a_{n}\left(  \mathrm{out}\right)  \left\vert 0,\mathrm{out}\right\rangle
=b_{n}\left(  \mathrm{out}\right)  \left\vert 0,\mathrm{out}\right\rangle
=0,\;\forall n. \label{c9}%
\end{align}

Let us consider the magnetic momentum operator,%
\begin{equation}
\widehat{\mathcal{M}}_{\Omega}=\frac{\mu}{2}\int\left[  \Psi_{\Omega}\left(
t,\mathbf{r}\right)  ^{\dagger},\Psi_{\Omega}\left(  t,\mathbf{r}\right)
\right]  _{-}d\mathbf{r} \label{c10.1}%
\end{equation}
and the operator of the kinetic energy of the quantum Dirac field
$\Psi_{\Omega}\left(  t,\mathbf{r}\right)  $ in the domain $\Omega$,%
\begin{equation}
\widehat{\mathcal{H}}_{\Omega}^{kin}=\int\Psi_{\Omega}\left(  t,\mathbf{r}%
\right)  ^{\dagger}\left[  \hat{\Pi}_{z}+\mu B_{z}\left(  y\right)  \right]
\sqrt{1+\left[  \frac{\hat{p}^{3}}{\hat{\Pi}_{z}+\mu B_{z}\left(  y\right)
}\right]  ^{2}}\Psi_{\Omega}\left(  t,\mathbf{r}\right)  d\mathbf{r-}%
H_{\Omega}^{0}\mathbf{\ } \label{c10.2}%
\end{equation}
where $\Psi_{\Omega}\left(  t,\mathbf{r}\right)  =\sum_{n\in\Omega}\Psi
_{n}\left(  t,\mathbf{r}\right)  $ and $H_{\Omega}^{0}=\left\langle
0,\mathrm{in}\left\vert \widehat{\mathcal{H}}_{\Omega}^{kin}\right\vert
0,\mathrm{in}\right\rangle $ is the constant term corresponding to the energy
of vacuum fluctuations. Using relations (\ref{i10})-(\ref{i12}), (\ref{rel1}),
and (\ref{UR1}), one can represent these operators in equivalent diagonal
forms as follows
\begin{align}
\widehat{\mathcal{M}}_{\Omega}  &  =\mu\sum_{n\in\Omega}\left[  a_{n}^{\dag
}(\mathrm{in})a_{n}(\mathrm{in})-b_{n}^{\dag}(\mathrm{in})b_{n}(\mathrm{in}%
)\right] \nonumber\\
&  =\mu\sum_{n\in\Omega}\left[  a_{n}^{\dag}(\mathrm{out})a_{n}(\mathrm{out}%
)-b_{n}^{\dag}(\mathrm{out})b_{n}(\mathrm{out})\right]  ;\nonumber\\
\widehat{\mathcal{H}}_{\Omega}^{kin}  &  =\sum_{n\in\Omega}\left[
\ ^{-}\mathcal{E}_{n}a_{n}^{\dag}(\mathrm{in})a_{n}(\mathrm{in})-\ _{-}%
\mathcal{E}_{n}b_{n}^{\dag}(\mathrm{in})b_{n}(\mathrm{in})\right] \nonumber\\
&  =\sum_{n\in\Omega}\left[  \ ^{+}\mathcal{E}_{n}a_{n}^{\dag}(\mathrm{out}%
)a_{n}(\mathrm{out})-\ _{+}\mathcal{E}_{n}b_{n}^{\dag}(\mathrm{out}%
)b_{n}(\mathrm{out})\right]  , \label{c10.3}%
\end{align}
where%
\[
_{\zeta}\mathcal{E}_{n}\mathcal{=}E_{+1}\left(  R\right)  +\frac{1}%
{2}U\left\vert g\left(  _{+}\left\vert ^{-}\right.  \right)  \right\vert
^{-2},\;\;^{\zeta}\mathcal{E}_{n}=E_{+1}\left(  L\right)  -\frac{1}%
{2}U\left\vert g\left(  _{+}\left\vert ^{-}\right.  \right)  \right\vert
^{-2},
\]
see details in our forthcoming work \cite{GavGit12}. We suppose that%
\begin{equation}
^{\zeta}\mathcal{E}_{n}>0,\;\;_{\zeta}\mathcal{E}_{n}<0, \label{i20}%
\end{equation}
in the external field under consideration, so that the signs of the energies
$^{\zeta}\mathcal{E}_{n}$ and $_{\zeta}\mathcal{E}_{n}$ are determined by the
signs of $\pi_{+1}\left(  R/L\right)  $. In known solvable cases the
inequalities (\ref{i20}) hold true; for example, see Refs.
\cite{Nik-barrier-70,Nikis79,Nik04}. Thus, the operator $\widehat{\mathcal{H}%
}_{\Omega}^{kin}$ is positively defined. This fact provides a consistent
quantization in terms of particles and antiparticles in the range $\Omega$.

Kinetic energy must be positive for any wave packets of both particles and
antiparticles. This is why particle wave packets are situated in the region
$S_{L}$ and antiparticle wave packets are situated in the region $S_{R}$, that
is, there is a total reflection from $S_{\mathrm{int}}$ for both particles and
antiparticles. This is consistent with the physical meaning. Note that the
expressions $\left(  ^{\zeta}\psi_{n},^{\zeta^{\prime}}\psi_{n^{\prime}%
}\right)  _{x}$ and $\left(  -1\right)  \left(  _{\zeta}\psi_{n}%
,_{\zeta^{\prime}}\psi_{n^{\prime}}\right)  _{x}$, given by Eq. (\ref{g13}),
are the probability currents of particles and antiparticles through the
surface $y=\mathrm{const}$, respectively. The particle and antiparticle
currents are positive for $\zeta=-1$ and negative for $\zeta=+1$. Thus, we see
that for $s=+1$\ the functions $^{+}\psi_{n}\left(  t,\mathbf{r}\right)  $ and
$_{+}\psi_{n}\left(  t,\mathbf{r}\right)  $ describe \ outgoing particles and
antiparticles, while the functions $^{-}\psi_{n}\left(  t,\mathbf{r}\right)  $
and $_{-}\psi_{n}\left(  t,\mathbf{r}\right)  $ describe incoming particles
and antiparticles, respectively. The\textrm{\ }particle-antiparticle and
causal identification of the wave functions (\ref{g5}) is unique in the
framework of QFT.

The vacuum corresponds to the absence of incoming particles and antiparticles.
In such a case the presence of outgoing particles and antiparticles indicates
particle creation from the vacuum. The effect of particle creation implies
constant currents of outgoing particles and antiparticles. These currents are
equal in the regions $S_{L}$ and $S_{R}$.

Then taking into account Eqs. (\ref{c5}) and (\ref{c6}), we obtain direct and
inverse linear canonical transformations between the "$\mathrm{in}$" and
"$\mathrm{out}$" creation and annihilation operators (Bogolyubov
transformations):
\begin{align}
&  a_{n}\left(  \mathrm{out}\right)  =g\left(  ^{-}\left\vert _{+}\right.
\right)  ^{-1}g\left(  ^{+}\left\vert _{+}\right.  \right)  a_{n}\left(
\mathrm{in}\right)  -g\left(  _{-}\left\vert ^{+}\right.  \right)  ^{-1}%
b_{n}^{\dagger}\left(  \mathrm{in}\right)  ,\nonumber\\
&  b_{n}^{\dagger}(\mathrm{out})=g\left(  ^{-}\left\vert _{+}\right.  \right)
^{-1}a_{n}\left(  \mathrm{in}\right)  +g\left(  _{-}\left\vert ^{+}\right.
\right)  ^{-1}g\left(  _{+}\left\vert ^{+}\right.  \right)  b_{n}^{\dagger
}(\mathrm{in});\nonumber\\
&  a_{n}\left(  \mathrm{in}\right)  =g\left(  ^{+}\left\vert _{-}\right.
\right)  ^{-1}g\left(  ^{-}\left\vert _{-}\right.  \right)  a_{n}\left(
\mathrm{out}\right)  +g\left(  _{+}\left\vert ^{-}\right.  \right)  ^{-1}%
b_{n}^{\dagger}(\mathrm{out}).\nonumber\\
&  b_{n}^{\dagger}(\mathrm{in})=-g\left(  ^{+}\left\vert _{-}\right.  \right)
^{-1}a_{n}\left(  \mathrm{out}\right)  +g\left(  _{+}\left\vert ^{-}\right.
\right)  ^{-1}g\left(  _{-}\left\vert ^{-}\right.  \right)  b_{n}^{\dagger
}(\mathrm{out}). \label{cq17}%
\end{align}
These transformations are similar to that used by Nikishov in the problem of
charged-particle scattering on an electric step \cite{Nik-barrier-70,Nikis79}.

With the help of the transformations (\ref{cq17}), we calculate the
differential mean number of created particles and antiparticles%
\begin{align}
N_{n}^{(+)}  &  =\left\langle 0,\mathrm{in}\right\vert a_{n}^{\dag}\left(
\mathrm{out}\right)  a_{n}\left(  \mathrm{out}\right)  \left\vert
0,\mathrm{in}\right\rangle =\left\vert g\left(  _{-}\left\vert ^{+}\right.
\right)  \right\vert ^{-2},\nonumber\\
N_{n}^{(-)}  &  =\left\langle 0,\mathrm{in}\right\vert b_{n}^{\dag}\left(
\mathrm{out}\right)  b_{n}\left(  \mathrm{out}\right)  \left\vert
0,\mathrm{in}\right\rangle =\left\vert g\left(  _{+}\left\vert ^{-}\right.
\right)  \right\vert ^{-2}. \label{c18}%
\end{align}
The relations (\ref{UR2}) imply the equality%
\[
N_{n}^{(+)}=N_{n}^{(-)}=N_{n},
\]
which allows us to treat $N_{n}$ as the differential mean number of created
pairs. The total number $N$ of created pairs is the sum%
\begin{equation}
N=\sum_{n\in\Omega}N_{n}. \label{TN}%
\end{equation}

The elementary relative probability amplitudes of particle creation,
annihilation, and scattering are defined as follows%

\begin{align}
&  c_{v}=\langle0,\mathrm{out}|0,\mathrm{in}\rangle\,,\nonumber\\
&  w\left(  +|+\right)  _{n^{\prime}n}=c_{v}^{-1}\langle0,\mathrm{out}%
\left\vert a_{n^{\prime}}\left(  \mathrm{out}\right)  a_{n}^{\dagger
}(\mathrm{in})\right\vert 0,\mathrm{in}\rangle,\nonumber\\
&  w\left(  -|-\right)  _{nn^{\prime}}=c_{v}^{-1}\langle0,\mathrm{out}%
\left\vert b_{n^{\prime}}\left(  \mathrm{out}\right)  b_{n}^{\dagger
}(\mathrm{in})\right\vert 0,\mathrm{in}\rangle\,,\nonumber\\
&  w\left(  0|-+\right)  _{nn^{\prime}}=c_{v}^{-1}\langle0,\mathrm{out}%
\left\vert b_{n}^{\dagger}(\mathrm{in})a_{n^{\prime}}^{\dagger}(\mathrm{in}%
)\right\vert 0,\mathrm{in}\rangle\,,\nonumber\\
&  w\left(  +-|0\right)  _{n^{\prime}n}=c_{v}^{-1}\langle0,\mathrm{out}%
\left\vert a_{n^{\prime}}\left(  \mathrm{out}\right)  b_{n}\left(
\mathrm{out}\right)  \right\vert 0,\mathrm{in}\rangle\,, \label{cq22}%
\end{align}
where $c_{v}$ is the vacuum-to-vacuum transition amplitude. One can see that
the amplitudes (\ref{cq22}) are diagonal%
\begin{align}
&  w\left(  +|+\right)  _{n^{\prime}n}=\delta_{n,n^{\prime}}w_{n}\left(
+|+\right)  ,\;w\left(  -|-\right)  _{nn^{\prime}}=\delta_{n,n^{\prime}}%
w_{n}\left(  -|-\right)  ,\nonumber\\
&  w\left(  0|-+\right)  _{nn^{\prime}}=\delta_{n,n^{\prime}}w_{n}\left(
0|-+\right)  ,\;w\left(  +-|0\right)  _{n^{\prime}n}=\delta_{n,n^{\prime}%
}w_{n}\left(  +-|0\right)  , \label{cq24}%
\end{align}
and can be expressed via the coefficients $g\left(  ^{\zeta^{\prime}%
}\left\vert _{\zeta}\right.  \right)  $ as follows:%
\begin{align}
&  w_{n}\left(  +|+\right)  =g\left(  ^{+}\left\vert _{-}\right.  \right)
g\left(  ^{-}\left\vert _{-}\right.  \right)  ^{-1}=g\left(  _{+}\left\vert
^{-}\right.  \right)  g\left(  _{+}\left\vert ^{+}\right.  \right)
^{-1},\nonumber\\
&  w_{n}\left(  -|-\right)  =g\left(  ^{-}\left\vert _{+}\right.  \right)
g\left(  ^{-}\left\vert _{-}\right.  \right)  ^{-1}=g\left(  _{-}\left\vert
^{+}\right.  \right)  g\left(  _{+}\left\vert ^{+}\right.  \right)
^{-1},\nonumber\\
&  w_{n}\left(  +-|0\right)  =g\left(  _{+}\left\vert ^{+}\right.  \right)
^{-1},\;w_{n}\left(  0|-+\right)  =-g\left(  ^{-}\left\vert _{-}\right.
\right)  ^{-1}, \label{cq25}%
\end{align}
where the transformations (\ref{cq17}) are used.

One can express the probabilities of particle scattering and pair creation for
quantum numbers $n\in\Omega$ and the probability for the vacuum to remain a
vacuum via the differential mean numbers $N_{n}$ as follows
\begin{align}
&  P(+|+)_{nn^{\prime}}=|<0,\mathrm{out}|a_{n}(\mathrm{out})a_{n^{\prime}%
}^{\dagger}(in)|0,\mathrm{in}>|^{2}=\delta_{n,n^{\prime}}\frac{1}{1-N_{n}%
}P_{v}\;,\nonumber\\
&  P(-+|0)_{nn^{\prime}}=|<0,\mathrm{out}|b_{n}(\mathrm{out})a_{n^{\prime}%
}(\mathrm{out})|0,\mathrm{in}>|^{2}=\delta_{n,n^{\prime}}\frac{N_{n}}{1-N_{n}%
}P_{v}\;,\nonumber\\
&  P_{v}=|c_{v}|^{2}=\exp\left\{  \sum_{n\in\Omega}\ln\left(  1-N_{n}\right)
\right\}  , \label{c26}%
\end{align}
see details in our forthcoming work \cite{GavGit12}. The probabilities for the
antiparticle scattering and the pair annihilation are described by the same
expressions $P(+|+)$ and $P(-+|0)$, respectively.

\section{Quasilinear magnetic field\label{S4}}

Here, we consider a specific case of an inhomogeneous magnetic field, namely a
field linearly growing on an interval $L_{y}.$ More exactly, the field has the
form%
\[
B_{z}(y)=\left\{
\begin{array}
[c]{l}%
B_{0},\ y<0\\
B_{0}+B^{\prime}y,\ y\in\lbrack0,L_{y}]\\
B_{0}+B^{\prime}L_{y},\ y>L_{y}%
\end{array}
\right.  ,
\]
where $B^{\prime}>0$ and $B_{0}=-$ $B^{\prime}L_{y}/2$. Let us call such a
field a quasilinear magnetic field. Consider the case given by the condition
\begin{equation}
\sqrt{\left\vert \mu B^{\prime}\right\vert }L_{y}\gg\max\left\{
1,m/\sqrt{\left\vert \mu B^{\prime}\right\vert }\right\}  , \label{SC}%
\end{equation}
which implies that there is particle creation in a wide enough range $\Omega$
of momenta given by condition (\ref{g14}). One can demonstrate, similar to the
case considered in Ref. \cite{GavG96} (see also Ref. \cite{WongW88}), that
leading contributions to the differential mean numbers $N_{n}$ of created
pairs do not depend on $L_{y}$ in the limit $L_{y}\rightarrow\infty$. This is
why, it is enough to consider the case of linearly growing magnetic field.
Equation (\ref{g9}) in the latter field for the function $\varphi_{n,\chi
}\left(  y\right)  $ given by Eq. (\ref{g6c}) can be written as%
\begin{align}
&  \left(  \frac{d^{2}}{d\xi^{2}}+\xi^{2}-\lambda+i\chi\right)  \varphi
_{n,\chi}\left(  y\right)  =0,\nonumber\\
&  \xi=\sqrt{\left\vert \mu\right\vert B^{\prime}}\left[  y+\left(  \left\vert
\mu\right\vert B^{\prime}\right)  ^{-1}\left(  \left\vert \mu\right\vert
B_{0}-\omega\right)  \right]  ,\ \lambda=\frac{m^{2}+p_{x}^{2}}{\left\vert \mu
B^{\prime}\right\vert }. \label{l1}%
\end{align}
Solutions of this equation, obeying the boundary conditions (\ref{g7a}) and
(\ref{g7b}), have the form%
\begin{equation}
{}_{+}^{-}\varphi_{n,\chi}\left(  y\right)  =\mathcal{N}_{\chi}D_{-\nu
-1}\left[  \pm(1+i)\xi\right]  ,\;\;_{-}^{+}\varphi_{n,\chi}\left(  y\right)
=\mathcal{N}_{\chi}D_{\nu}\left[  \pm(1-i)\xi\right]  , \label{l2}%
\end{equation}
where $D_{\nu}\left(  z\right)  $ are Weber parabolic cylinder (WPC)
functions, $\nu=-\left(  i\lambda+1+\chi\right)  /2.$ With the help of an
asymptotic expansion of WPC functions, one can verify the validity of the
boundary conditions (\ref{g7a}) and (\ref{g7b}). Using the solutions
(\ref{l2}), we construct the sets $\left\{  _{\zeta}\psi_{n}\left(
t,\mathbf{r}\right)  \right\}  $ and $\left\{  ^{\zeta}\psi_{n}\left(
t,\mathbf{r}\right)  \right\}  $ of solutions of the Dirac-Pauli equation.

The obtained form of solutions formally coincide with the one found in Refs.
\cite{Nik-barrier-70,Nikis79,WongW88} for the case of charged-particle
creation by a constant uniform electric field (compare with Ref.
\cite{GavG96}). Note that our identification of wave functions is in agreement
with one given by Nikishov for such a special case. This allows us to use
these calculations to find differential mean numbers of created pairs given by
Eq. (\ref{c18}).

In the limit $\sqrt{\left\vert \mu B^{\prime}\right\vert }L_{y}\gg K,\ $where
$K$ is a given arbitrary number $K\gg\max\left\{  1,m/\sqrt{\left\vert \mu
B^{\prime}\right\vert }\right\}  ,$ and if $\omega$ and $p_{z}$ satisfy the
condition
\[
\left\vert \omega\right\vert <\omega_{\max},\;\left\vert p_{z}\right\vert
<\omega_{\max},\;\omega_{\max}=\left\vert \mu B^{\prime}\right\vert
L_{y}/2-\sqrt{\left\vert \mu B^{\prime}\right\vert }K,
\]
we obtain%
\begin{equation}
N_{n}=e^{-\pi\lambda} \label{l3}%
\end{equation}

Following the idea of finite work regularization presented in Ref.
\cite{GavG96}, one can show that an exact expression for $N_{n}$ is rapidly
decreasing as $\left\vert \omega\right\vert \rightarrow\infty$ due to the
finite work of this field, $\left\vert \mu B^{\prime}\right\vert L_{y}$,
\ that is, $\omega_{\max}$ is an effective maximum value of the quantum number
$\left\vert \omega\right\vert $ for the quasilinear field under consideration.
The maximum value for $\left\vert p_{z}\right\vert $ from the range $\Omega$
follows from condition (\ref{g14}). One can check that the mean numbers do not
depend on the sign of $\mu B^{\prime}$ and on the spin polarization $s$. Note,
however, that unlike the case of particle creation due to the electric
potential step, the neutral particles (antiparticles) created with different
$s$ form fluxes aimed in opposite directions. The leading approximation given
by expression (\ref{l3}) does not depend on the quantum numbers $\omega$ and
$p_{z}$. Although the result (\ref{l3}) has been derived for $B^{\prime
}=\mathrm{const}$ field, it can be applicable to a spatially slowly varying
$B^{\prime}\left(  y\right)  $ as a good approximation if its gradient
variation is sufficiently small in comparison with the mean value
$\overline{B^{\prime}}$ on the interval $\left[  -L_{y}/2,L_{y}/2\right]  $,
$\overline{B^{\prime}}^{-1}\partial_{y}B^{\prime}\left(  y\right)  L_{y}\ll1$.

Let us calculate the total number $\mathcal{N}_{s}$ of created pairs with
given $s$ defined by Eq. (\ref{TN}). To do this we go over from the sum to an
integral,%
\[
\sum_{p_{x},p_{z},p_{0}}\left(  \cdots\right)  \Longrightarrow\frac{L_{x}%
L_{z}T}{\left(  2\pi\right)  ^{3}}\int\left(  \cdots\right)  dp_{x}%
dp_{z}dp_{0}\;.
\]
Taking into account that the exact distribution $N_{n}$ plays the role of a
cutoff factor in the integral over $\omega$, $p_{x}$, and $p_{z}$ we represent
the total number $\mathcal{N}_{s}$ in the form%
\begin{equation}
\mathcal{N}_{s}=2\int_{0}^{\omega_{\max}}dp_{z}\mathcal{N}_{s,p_{z}%
},\;\;\mathcal{N}_{s,p_{z}}=\frac{L_{x}L_{z}T}{\left(  2\pi\right)  ^{3}}\int
dp_{x}\int_{0}^{\omega_{\max}^{2}}\frac{N_{n}d\omega^{2}}{\sqrt{\omega
^{2}+p_{z}^{2}}}, \label{l4}%
\end{equation}
where the relation $p_{0}=\omega\sqrt{1+\left(  p_{z}/\omega\right)  ^{2}}$
from Eq. (\ref{g4a}) is used. We obtain the leading contribution in Eq.
(\ref{l4}) as follows%
\begin{equation}
\mathcal{N}_{s,p_{z}}=L_{x}L_{z}Tn_{s,p_{z}},\;\;n_{s,p_{z}}=\frac
{\sqrt{\left\vert \mu B^{\prime}\right\vert }}{4\pi^{3}}\exp\left(  -\frac{\pi
m^{2}}{\left\vert \mu B^{\prime}\right\vert }\right)  \left(  \sqrt
{\omega_{\max}^{2}+p_{z}^{2}}-\left\vert p_{z}\right\vert \right)  .
\label{15}%
\end{equation}
From Eq. (\ref{15}), we see that the leading term of the density $n_{s,p_{z}}$
is linear function of the length $L_{y}$ for sufficiently small momentum
$p_{z}$, $\left\vert p_{z}\right\vert \ll\omega_{\max}$, that is, the density
of the particles created per unit spacetime volume, $n_{s,p_{z}}/L_{y},$ is
uniform. Of course, this is not the case when $\left\vert p_{z}\right\vert $
is not small. Thus, we see a complete similarity between the case of particle
creation due to a quasiuniform electric field and a quasilinear magnetic field
for small momenta $p_{z}$ only. Using Eq. (\ref{l4}), we obtain the total
number $\mathcal{N}_{s}$ of created pairs with a given $s$ in the form%
\begin{equation}
\mathcal{N}_{s}=\frac{\sqrt{2}-1+\ln\left(  1+\sqrt{2}\right)  }{16\pi^{3}%
}TL_{x}L_{z}L_{y}^{2}\left\vert \mu B^{\prime}\right\vert ^{5/2}\exp\left(
-\frac{\pi m^{2}}{\left\vert \mu B^{\prime}\right\vert }\right)  . \label{l6}%
\end{equation}
The total number of created pairs with both $s=\pm1$ is $\mathcal{N=}%
N_{+1}+N_{-1}$.

The vacuum-to-vacuum transition probability defined in Eq. (\ref{c26}) can be
calculated in the same way. Then we express it via the total number $N$ as
follows%
\begin{equation}
P_{v}=\exp\left(  -\beta N\right)  ,\;\;\beta=\sum_{l=0}^{\infty}\left(
l+1\right)  ^{-3/2}\exp\left(  -\frac{l\pi m^{2}}{\left\vert \mu B^{\prime
}\right\vert }\right)  . \label{l7}%
\end{equation}

\section{Discussion\label{S5}}

It should be noted that the particle creation in the linearly growing magnetic
field represents a wide class of physical situations where the gradient of
magnetic fields is slowly varying in big enough but restricted areas.\emph{\ }%
One can also see that\ the leading contribution to differential mean numbers
of created pairs\ in such fields does not depend on the asymptotic behavior of
the magnetic field as the size of the heterogeneity tends to infinity. This
allows one to make some general conclusions from the obtained results.

First of all in $3+1$ dimensions, both the total number $N$ of created pairs
and the vacuum-to-vacuum transition probability $P_{v}$ given by Eqs.
(\ref{l6}) and (\ref{l7}), respectively depend only on the gradient of the
magnetic field, but not on its strength, similarly to what happens in $2+1$
dimensions \cite{Lin99}. Both quantities are finite for the finite spacetime
volume of field inhomogeneity. In particular, it seams that the level crossing
discovered in Ref. \cite{LeeY06} for the system of neutral fermions
interacting with strong uniform magnetic field due to an anomalous magnetic
moment is a result of improper treatment of the weak-field case spectrum. The
arbitrarily strong uniform magnetic field is stable with respect to the
creation of neutral fermions with anomalous magnetic moment and this fact does
not depend on the spacetime dimension.

Secondly, due to the nonperturbative consideration in the framework of QFT,
some results could emerge that can be difficult to expect when remaining in
the framework of one-particle quantum mechanics. In particular, in the case
under consideration of neutral particle creation, we have to stress the
following nontrivial peculiarities.

a) In contrast to the case of charged particles that are accelerated by an
electric field in directions that are defined by their charges, both the
neutral particles and antiparticles with opposite values of the conserved spin
polarization $s$\ have---due to the Pauli interaction---opposite directions of
acceleration. For this reason, only states with a definite $s$\ are
localizable and can form wave packets in the asymptotic regions. In fact, in
the problem under consideration, it is convenient to speak about two different
species of particles and antiparticles that are labeled by the sign of
$s$\textbf{. }For each kind $s$ there exist "\textrm{in}" and "\textrm{out}"
sets of solutions of the Dirac-Pauli equation\textbf{\ }that in QFT define the
corresponding "\textrm{in}" and "\textrm{out}" states\textbf{. }Note that
neutral particles (antiparticles) that are created by the external field with
different spin polarization $s$ form fluxes directed in opposite directions.
In a sense this explains the fact that quantization in terms of neutral
particles and antiparticles in $d\geq3+1$ dimensions is possible only in terms
of exact solutions with definite spin integrals of motion [in the case under
consideration, this integral of motion is the operator $\hat{R}$ given by Eq.
(\ref{g2b})]. This means that in models with a nonminimal interaction with an
external field and with $d\geq3+1$ the formal second quantization similar to
QED may not work.

b)\emph{\ }At a certain stage, calculations of the creation of neutral
fermions from the vacuum by inhomogeneous magnetic fields are technically
reduced to the calculations of the creation of charge particles from the
vacuum by corresponding electric fields. This allowed us to use some technical
results obtained earlier in QED regardling charged Dirac particles. However,
this does not mean that physically both effects are similar. For example---in
contrast to the case of charged-particle creation in a constant electric
field---in the case of the neutral fermion creation, the total number $N$ of
created pairs and $\ln P_{v}^{-1}$ are not linear in all length scales of \ an
accelerating field.{\Huge \ }This peculiarity is due to the different form of
the area in the phase space where particle creation occurs.

It is known that the Schwinger method of an effective action \cite{S51} is
convenient for semiclassical calculations of pair creation from vacuum due to
an electric-like field \cite{RufVSh10}. In this approach, one calculates the
probability for the vacuum to remain a vacuum using the following Schwinger
representation
\begin{equation}
P_{v}=e^{-2\operatorname{Im}W}, \label{d1}%
\end{equation}
where $W$ is the one-loop effective action of the corresponding QFT model. The
worldline approaches to QED are suitable for realistic backgrounds
\cite{worldline-inst11,worldline-num05}. In particular, for the case of the
creation of neutral fermions with an anomalous magnetic moment, representation
(\ref{d1}) was used in Refs. \cite{Lin99,LeeY06}. One can find a relation
between our results---obtained in the framework of canonically quantized field
theory---and the latter approach. To this end we present the quantity
(\ref{d1}) as an infinite product,
\begin{equation}
P_{v}=\prod_{n\in\Omega}e^{-2\operatorname{Im}W_{n}}\,, \label{d2}%
\end{equation}
where the quantum numbers $n=(p_{x},p_{z},\omega,s)\in\Omega$ (eigenvalues of
the corresponding integrals of motion) are used for parametrization, so that
the effective action $W$ is written as a sum $W=\sum_{n}W_{n}$. Then,
$e^{-2\operatorname{Im}W_{n}}$ is the vacuum-persistence probability in a cell
of the space of quantum numbers $n$. Using an exact expression for $P_{v}$ in
terms of the differential mean values $N_{n}$, given by Eq. (\ref{c26}), we
obtain the following relation
\begin{equation}
2\operatorname{Im}W_{n}=-\ln\left(  1-N_{n}\right)  \,. \label{d3}%
\end{equation}
As was noted above, the creation of neutral fermions with given quantum
numbers $n$ is reduced to the problem of charged-particle creation from vacuum
by a corresponding electric step. Then relation (\ref{d3}), well known for the
case of a constant electric field \cite{Nik-barrier-70,Nikis79}, also holds
for the creation of neutral fermions in a linearly growing magnetic field.
This means that the Schwinger method works for the case under consideration,
provided we have a suitable parameterization. However, we see that the total
quantities $N$ (and $\ln P_{v}^{-1}$) in $3+1$ dimensions are quadratic in
$L_{y}$. This is a consequence of the fact that the number of states with all
possible $\omega$ and $p_{z}$ excited by the field $B^{\prime}$ is quadratic
in the kinetic momentum $\left\vert \mu B^{\prime}\right\vert L_{y}$. This is
also the reason why the density of created pairs and the density of
$\operatorname{Im}W$ per unit of length $L_{y}$ are not constant. In this case
the divergence of the effective action $W$ as $L_{y}\rightarrow\infty$ is not
linear and it is quite difficult to invent a reliable method of regularization
of $W$ for a linearly growing magnetic field in the framework of the Schwinger
approach, if the parametrization is not appropriately chosen, as was done
above. We believe that ignoring this fact was the main cause of the
questionable results in Ref. \cite{LeeY06}. On the other hand, in $2+1$
dimensions, there is only one spin polarization and the integration over
$p_{z}$ is absent, that is, the calculation of the quantities $N$ (and $\ln
P_{v}^{-1}$) for created neutral fermions by a linearly growing magnetic field
is completely reduced to the problem of charged-particle creation from vacuum
by a constant electric field. Then the expression for $P_{v}$ obtained in Ref.
\cite{Lin99} is in agreement with our result for $N_{n}$, given by Eq.
(\ref{l3}). Note that our techniques in the framework of QFT can be used to
separate the divergent term of $\operatorname{Im}W$ as $L_{y}\rightarrow
\infty$ in the framework of the effective action techniques and to relate it
to pair creation, cf. Ref. \cite{GG08}. It means that recent computational
developments \cite{worldline-inst11,worldline-num05} can also be extended to
calculate the effects of particle creation with an anomalous magnetic moment.

The cases with opposite values of the spin polarization $s$ differ only in
that they have opposite directions of all the motions and all the asymptotic
regions with respect to a nonzero-gradient region of the magnetic field. Then,
the neutral particles (antiparticles) created with different $s$ form fluxes
that are moving in opposite directions. The probabilities of all the processes
are equal for different values of $s$. We see that the created flux aimed in
one of the directions is formed from fluxes of particles and antiparticles of
equal intensity and with the same magnetic moments parallel to the external
field. In such a flux particle and antiparticle velocities that are
perpendicular to the plane of the magnetic moment and flux direction are
essentially depressed. This is a typical property of neutral fermions created
by inhomogeneous magnetic fields that can be used to observe their effects in
astrophysical situations.

As follows from the obtained results, the effective creation of neutral
fermions from vacuum starts when there exists a big enough difference between
the asymptotic magnetic fields, i.e., $U>2m$. Let us suppose that the magnetic
field under consideration achieves its maximal value $\left\vert B_{\max
}\right\vert $ inside of a finite region and is absent outside this region. In
this case, the minimal value of the quantity $\left\vert B_{\max}\right\vert $
which provides the effective particle creation is $\left\vert B_{\max
}\right\vert \sim B_{cr}=2m/\left\vert \mu\right\vert $. It is convenient to
express the magnetic moment $\mu$ in terms of the Bohr magneton, $\left\vert
\mu\right\vert =2c_{\mu}$ $\mu_{B}$, $\mu_{B}=\frac{e}{2m_{e}}$, and the
particle mass $m$ in terms of the electron mass $m_{e}$, $m=c_{m}m_{e}$, such
that $c_{\mu}$ and $c_{m}$ are the corresponding dimensionless quantities.
Then the characteristic magnetic field $B_{cr}$ in the problem under
consideration is
\[
B_{cr}=2B^{QED}\frac{c_{m}}{c_{\mu}},\;B_{QED}=m_{e}^{2}/e=m_{e}^{2}%
c^{3}/e\hbar\simeq4.4\cdot10^{13}G,
\]
where $B_{QED}$ is the characteristic magnetic field value above which the
nonlinearity of QED becomes actual. There are two species of neutral fermions
among the known elementary particles: the neutron and the active neutrino. For
the neutrons $c_{m}/c_{\mu}\sim10^{6}$ which implies $B_{cr}^{(n)}\sim
10^{20}\;\mathrm{G}$. In the active neutrino case the optimistic estimation is
$c_{\mu}\sim10^{-12}$. Cosmological constraints \textbf{\ }indicate that the
total active neutrino mass is below $0.3\;\mathrm{eV}$ \cite{nu-mass-cosm12}.
Then supposing that the mass of the active neutrino is of the order $m_{\nu
}\sim0.1\;\mathrm{eV}$, i.e., $c_{m}\sim10^{-7}$, we obtain $\sim10^{5}$ for
the factor $c_{m}/c_{\mu}$, which implies that the critical value is
$B_{cr}^{(\nu)}\sim10^{19}\;\mathrm{G}$. However, it should be noted that if
the active neutrino mass is essentially less than $0.1\;\mathrm{eV}$ (which is
theoretically admissible) and, at the same time, its magnetic moment is not
significantly less than $10^{-12}\mu_{B}$, then it is possible that
$B_{cr}^{(\nu)}\ll10^{19}\;\mathrm{G}$. None of the neutrino models are
currently universally accepted, such that we do not have any theoretical
estimation of their masses and magnetic moments. We do not certainly know
whether neutrinos are Dirac or Majorana particles. Moreover, the neutrino
magnetic moment and therefore the ratio $c_{m}/c_{\mu}$ can depend on the
strength of a strong magnetic field; see, for example, Ref. \cite{MM-B02}.
This is why at present it is difficult to give more exact estimation for
$B_{cr}^{(\nu)}$.

Taking into account the possible existence of the light sterile neutrinos with
masses $M$ in the range of $1\;\mathrm{keV}$ \cite{Kuz09,LSnu12} and weak
observational constraints on their magnetic moment $\mu$
\cite{Sig-etal04,Gard09}, we propose the new scenario in which pairs of
sterile neutrinos and antineutrinos could be produced from their coupling to
an inhomogeneous magnetic field. For example, if $M=m_{e}/10$ then $\left\vert
\mu\right\vert <3.4\times10^{-5}\mu_{B}$ while if $M=m_{e}/100$ then
$\left\vert \mu\right\vert \lesssim10^{-4}\mu_{B}$ due to precision
electroweak measurements \cite{Gard09}. In the latter case, the most
optimistic estimation is $c_{m}/c_{\mu}\sim10^{2}$ which implies that the
critical value $B_{cr}^{(s\nu)}\sim10^{16}\;\mathrm{G}$. Sterile neutrinos
with masses of several keV are dark matter candidate. Thus, we have an
estimation of the critical value $B_{cr}^{(s\nu)}$ that is relevant for dark
matter. These constraints can be weakened by the mechanism of compositeness
and a variety of astrophysical constraints can be significantly weakened by
the candidate particle's mass. In this situation, one can use, for example,
the direct limits on $\left\vert \mu\right\vert $, which would follow from the
nonobservance of Faraday rotation at a given sensitivity, and see that
$\left\vert \mu\right\vert \lesssim\mu_{B}$ \ \cite{Gard09}. If $M=m_{e}/100$
then such a weak limit implies $B_{cr}^{(s\nu)}\sim10^{12}\;\mathrm{G}$.

One can see from the discussion presented in the Introduction that the
magnetic field in the magnetar cores made of quark matter can likely reach the
critical value $B_{cr}^{(n)}$ which is enough to create neutron-antineutron
pairs. Magnetic fields generated during a supernova explosion or in the
vicinity of magnetars are of the order $10^{15}-10^{16}\;\mathrm{G}$ or even
stronger, up to $10^{18}\;\mathrm{G}$. Such fields cannot create
neutron-antineutron pairs from the vacuum but are strong enough to create
neutrino-antineutrino pairs. In any case the vacuum instability with respect
to the creation of neutrinos and even neutrons in strong magnetic fields has
to be taken into account in the astrophysics. In particular, it may be of
significance for dark matter studies.

It follows from Eq. (\ref{l6}) that the intensity of fluxes of created pairs
turns out to be essential when the gradient $B^{\prime}$ is sufficiently
large, $\left\vert B^{\prime}\right\vert \sim\left\vert B_{\max}\right\vert
/L_{y}\sim m^{2}/\left\vert \mu\right\vert $, and the condition of
applicability of the model of the linearly growing magnetic field is valid,
$\sqrt{\left\vert \mu B^{\prime}\right\vert }L_{y}\gg1$. This implies the
following estimation for $\left\vert B_{\max}\right\vert $:
\[
\left\vert B_{\max}\right\vert \sim L_{y}mB_{cr},
\]
where $L_{y}m\gg1$. Thus, considering astrophysical objects, one has to take
into account the backreaction due to the vacuum instability in magnetic fields
with $\left\vert B_{\max}\right\vert \gg B_{cr}$. The magnetic moments of
created pairs are antiparallel in opposite asymptotic regions; the
corresponding induced magnetic field has a gradient that is opposite to the
gradient of the external magnetic field. Thus, neutral particle creation leads
to a smoothing of the initial magnetic field, which in turn prevents the
appearance of superstrong constant magnetic fields. In any case, background
magnetic fields greater than $B_{cr}^{(\nu)}$, $B_{cr}^{(s\nu)}$, and
$B_{cr}^{(n)}$ may create effects of the vacuum instability due to the above
considered mechanism. In particular, magnetic fields with $\left\vert B_{\max
}\right\vert \gg B_{cr}^{(s\nu)}$ can produce fluxes of pairs of sterile
neutrinos and antineutrinos, which could escape the star with an anisotropy
equal to the anisotropy in their production .

We hope that by applying similar approaches to quantum massive neutral
fermionic fields, interacting with external backgrounds \cite{DvoGit13}, we
can study the creation of Dirac and (probably) Majorana massive neutrinos from
the vacuum by an inhomogeneous background matter.

\subparagraph{{\protect\large Acknowledgements}}

We appreciate the anonymous referee for important remarks. The work of S.P.G.
is supported in part by FAPESP/Brasil and the program Bolsista CAPES/Brasil.
S.P.G. thanks University of São Paulo for hospitality. D.M.G. acknowledges the
permanent support of FAPESP and CNPq. S.P.G. and D.M.G. thank for partial
support by the Grant No. 14.B37.21.0911 of Russian Ministry of Science.

\end{document}